%% file: main.tex
\documentclass[conference]{IEEEtran}

\input{setup.tex}
\IEEEoverridecommandlockouts

\begin{document}
\title{The Capacity of Multi-user Private Information Retrieval for Computationally Limited Databases}

\author{\IEEEauthorblockN{William Barnhart\IEEEauthorrefmark{1}
 and
Zhi Tian\IEEEauthorrefmark{2}}
\IEEEauthorblockA{Department of Electrical and Computer Engineering,
George Mason University\\
Fairfax, VA, USA\\
\IEEEauthorrefmark{1}wbarnha2@gmu.edu,
\IEEEauthorrefmark{2}ztian1@gmu.edu}
\thanks{This work was partly supported by the US NSF grant \# SaTC-1704274.}
}



\maketitle
\begin{abstract}
We present a private information retrieval (PIR) scheme that allows a user to retrieve a single message from an arbitrary number of databases by colluding with other users while hiding the desired message index. This scheme is of particular significance when there is only one accessible database- a special case that turns out to be more challenging for PIR in the multi-database case. The upper bound for privacy-preserving capacity for these scenarios is $C=(1+\frac{1}{S}+\cdots+\frac{1}{S^{K-1}})^{-1}$, where $K$ is the number of messages and $S$ represents the quantity of information sources such as $S=N+U-1$ for $U$ users and $N$ databases. We show that the proposed information retrieval scheme attains the capacity bound even when only one database is present, which differs from most existing works that hinge on the access to multiple databases in order to hide user privacy. Unlike the multi-database case, this scheme capitalizes on the inability for a database to cross-reference queries made by multiple users due to computational complexity.
\end{abstract}

\vspace{9pt}

\section{Introduction}
The area of Private Information Retrieval (PIR) focuses on obtaining information from various databases while protecting user privacy. The goal of PIR is to maintain privacy while retrieving the largest amount of information possible while downloading the minimum quantity of irrelevant information necessary in order to obfuscate the desired message index. In the classical PIR setting \cite{chor_private_1998}, there are multiple non-communicating and replicated databases that contain the same messages from which a user may download a particular message of interest. When multiple identical databases are present, a user can download different segments of the desired message from different databases and then patch them into a complete message, thus hiding its intent for all databases. For this purpose, the user prepares a query for each database, such that the queries do not reveal the user's intent. From the perspective of information theory, the goal of query design is to maximize retrieval efficiency. Our motivation behind studying this problem in PIR stems from the problem of privacy-preserving information retrieval with a single database. We use capacity to determine how efficient we can create a PIR scheme in terms of the ratio of desired information bits to the total quantity of information amassed by a single user.

There has been considerable progress on topics that involve databases trying various strategies to compromise privacy, such as databases sharing information with other databases to compromise the privacy of a user, known as colluding databases \cite{sun_capacity_2018}. When up to $T$ of the $N$ databases are colluding, the setting is known as "T-private PIR" (TPIR). If all databases are colluding, then capacity for a single user is equivalent to the setting when accessing a single database, which is a major issue encountered in PIR. What has not been considered by \cite{sun_capacity_2016-1} is the presence of multiple users working together to obfuscate the message index while accessing information. This is known as "single-server multi-user PIR" (SSMUPIR), for which there is little prior work \cite{li_single-server_2018-1}. This solution can be expanded to solve other problems, such as systems with multiple databases. While \cite{li_single-server_2018-1} has successfully proven the feasibility for a multi-user setting, the maximum retrieval rate has yet to be found. There has been prior work for single-server PIR in \cite{kushilevitz_replication_1997} and becomes feasible when communication complexity satisfies a particular threshold. However, there has not yet been any systematic research on how a retrieval scheme can be designed for an individual user to interact with multiple users and an arbitrary number of databases while maintaining privacy.

\IEEEpubidadjcol

This paper aims to solve the capacity of both the single database problem and the multi-database problems. Our solution considers the unifying setting with an arbitrary number of databases because many real-world applications create collusion across networked databases. When instantiated to the little-studied single database setting, our novel approach derives from the observation that a setting with one database and multiple users can be mapped as a mirrored version of \cite{sun_capacity_2017}. This approach allows us to design a retrieval scheme and analyze maximum retrieval efficiency. We refer to the setting of multiple users accessing multiple databases as multi-user PIR (MUPIR). Our setting for introducing MUPIR stems from the basis for network coding in \cite{ahlswede_network_2000} to improve throughput via creating multiple channels. In our setting, we create more channels by increasing the number of users that can interact with a fixed set of databases. We focus on the problem created by databases examining queries made by a single user and determine that potential information sources, users and databases, should be treated as separate random variables. Along this line, \cite{yeung_new_1991} demonstrates how three random variables could be used for characterizing capacity, which could be useful for creating an extension of the work presented by \cite{sun_capacity_2017}. A query must be made in order to move information from a database to a user, which contains a linear combination of bits from multiple sources of information.

The ensuing paper is organized as follows. Section II defines the problem of interest. Section III presents the capacity of MUPIR, followed by a scheme that achieves this capacity. Section IV formally proves the capacity of MUPIR, and Section V demonstrates that the proposed scheme for MUPIR reaches the derived capacity. Concluding remarks are included in Section VI.

 \textit{Notation: To represent the set of answers $\left \{A^{[\theta]}_{1},\cdots,A^{[\theta]}_{U}\right \}$, $\theta$ being the desired message index and $U$ as the number of users, we use $A_{1:U}^{[\theta]}$. For $\left \{A^{[1]}_{1:U},\cdots,A^{[K]}_{1:U}\right \}$, the notation $ A^{[1:K]}_{1:U}$ is applied. To represent ${Q^{[\theta]}_{1:S}}$, we employ the compact form $\Q$. For any positive integer $A$, we denote an index set using $[A]=\left \{1,2,\cdots,A \right \}$. We utilize the notation Q $\sim$ A to indicate Q and A are identically distributed. Let $[U] \setminus u = [U']$, so that $u \notin [U']$.}

\section{Problem Formulation}
Consider a network of $U$ users who seek to access messages contained in $N$ databases. It is assumed that all of the $N$ databases contain the same $K$ messages $W_{k}$, $\forall{k} \in [K]$. Each message $W_{\theta}$ is made of $L$ bits, which can be downloaded bit by bit and then patched together to reconstruct the original message. A user seeks to retrieve message $W_{\theta}$ by having a user generate a total of $S$ queries for message $\theta$, denoted as $\p{Q}{\theta}{s}$, sent to all $S=N+U-1$ sources. Each source answers the original user with $A^{[\theta]}_{s}$, $\forall{s} \in [S]$, and allows for the desired message $W_{\theta}$ to be obtainable from a set of $D$ bits. The scheme utilized to retrieve information, $Q(s,\theta)$, generates $\p{Q}{\theta}{s}$. When query $\p{Q}{\theta}{s}$ is sent to user $u$, then the query is forwarded to a single, arbitrary database $n$. Answer $\p{A}{\theta}{s}$ is returned to user $u$, who returns the answer to the original user. If query $\p{Q}{\theta}{s}$ is sent to database $n$, then $\p{A}{\theta}{s}$ is directly returned to the original user.

Our objective is to retrieve a message $W_{\theta}$ from a database and other users without revealing the value of the message index, $\theta$, while obtaining the message at the maximum possible rate. We define rate as a ratio $R=L/D$, where $L$ is the length of the intended message $W_{\theta}$, and $D$ is total number of downloaded bits. In order for a single user to hide the message index from $U-1$ users and $N$ databases, all queries generated by $Q(s,\theta)$ must be identical in structure for the same sets of $K$ distinct messages. This leaves us to readapt the privacy constraint defined by [5, eq. (8)] as
\begin{equation}\label{privacy}
    (Q^{[1]}_{s},A^{[1]}_{s},W_{1:K})\sim (Q^{[k]}_{s},A^{[k]}_{s},W_{1:K}),\forall{s} \in [S], k \in [K].
\end{equation}
Additionally, to maintain privacy, each query $Q_{s}^{[\theta]}$, $\forall{s} \in [S]$ must be independent of the desired message index, $\theta$. We express this by evaluating the mutual information, $I(\cdot ; \cdot)$, between $Q_{1:S}^{[k]}$ and $W_{1:K}$, ensuring there is no definite overlap between the queries and messages. We formally express this as defined in [2, eq. (3)]:
\begin{equation}\label{independence}
    I(Q_{1:S}^{[k]};W_{1:K})=0,\quad\forall{k} \in [K].
\end{equation}
To represent the size of message $W_{\theta}$ in terms of bits, we use the binary entropy function $H(W_{\theta})$. We express the size of $W_{\theta}$ related to the size of other messages by using index $k$ in the form
\begin{equation}
    H(W_{\theta})=H(W_{k})=L,\quad\forall{k}\in [K].
\end{equation}
The purpose of finding capacity in PIR is to find the maximum rate of retrieval, which focuses on obtain the largest possible message by downloading the smallest amount of bits while satisfying the criterion for privacy. The total size (in bits) of all $K$ messages can be equivalently represented as
\begin{equation}
    \sum_{k=1}^{K}H(W_{k})=KH(W_{\theta})=H(W_{1:K}).
\end{equation}
 The total quantity of downloaded information by a central user from $S$ sources is defined as $D=SH(\p{A}{k}{1})$. Alternatively, for desired message $W_{\theta}$, $D=SH(\p{A}{\theta}{1})$ is downloaded. We are not concerned with the upload cost for sending queries, so we do not directly take into account the magnitude of $H(\p{Q}{\theta}{s})$. In order to decode the quantity information withheld in an answer, a user must have knowledge of all queries. Thus, we must evaluate the entropy of downloaded information in the form $H(\p{A}{\theta}{s}|\Q)$. We adapt the definition of rate in \cite{sun_capacity_2016-1}:
\begin{equation}\label{rate}
    R \triangleq \frac{L}{D} = \frac{H(W_{\theta})}{SH(A^{[\theta]}_{1}|\Q)}
\end{equation} 
We use the general notation $C_{\text{MUPIR}}$ to represent capacity in a multi-user and a multi-database setting. The relation between rate and capacity can be expressed as $R \leq C_{\text{MUPIR}}$, since it is not possible to achieve a more efficient rate than the capacity. In order to design an efficient retrieval scheme, \cite{sun_capacity_2017} states the following three principles, the first two of which are imposed to ensure user privacy, while the third principle is used to improve efficiency for retrieving information.\footnote{Each enumerated item corresponds to arrows indicating different steps of generating queries.}
\textit{\begin{itemize}[leftmargin=2.5em]
    \item[$(1)$]Queries to each information source obtain forced symmetry
    \item[$(2)$]Message symmetry is enforced within all answers and queries
    \item[$(3)$]Side information of undesired messages is exploited to retrieve new desired information
\end{itemize}}
Based upon these principles, we next analyze capacity and design retrieval algorithms for databases to address capacity-achieving solutions for classical PIR in order to evaluate the computational cost of thwarting our scheme. Our designs are developed for two scenarios: One is when the databases honestly respond to queries but do not analyze the queries, and the other is when databases can analyze and cross-reference queries from multiple users to determine the user's desired message index. In order to determine our scheme's usefulness, we must determine the computational cost associated with analyzing the query schemes. Evaluating Algorithms 1 and 2 of Appendices A and B, it is feasible to compute the desired message index with a computational complexity of $O(n^2)$. Due to the similarities between Algorithms 1 and 2, $n$ would have to correspond to the length of each query sent to a database. If a database does have the option to cross-reference queries, then the remaining set of $U-1$ would be forced to replicate all contents stored in a database to guarantee privacy. Given the scope of our problem, we do not take into consideration the efficiency of replicating all $K$ messages across all $U-1$ users.

\section{Main Results}
\subsection{PIR Capacity for Independent Databases}
We start with the first case assuming that users cannot collude and the databases are non-communicating. Under such assumptions, there has not been a great deal of research performed with multi-user PIR settings, specifically addressing multi-user and multi-database settings. We find that users can collude together against an arbitrary number of databases while satisfying the defined privacy constraints. Our main result on the user privacy-preserving capacity is given by Theorem 1.
\newtheorem{theorem}{Theorem}
\begin{theorem}
Our capacity for MUPIR applies to settings with $N$ non-curious or computationally-limited databases and $U$ trustworthy users. We characterize $S=N+U-1$ sources of information to have the capacity:
\begin{equation}\label{onencap}
    C_{\text{MUPIR}} =\bigg(1+\frac{1}{S}+\cdots+\frac{1}{S^{K-1}}\bigg)^{-1}
\end{equation}
\end{theorem}
Our theorem is similar in design to the theorem for classical PIR proposed in \cite{sun_capacity_2017}, however, our novel approach focuses on adding on additional users treated like databases in order to increase retrieval rate.

\subsection{Proposed Retrieval Scheme}
A functional system with multiple users and a single database can be generated by mirroring the classical PIR problem in \cite{sun_capacity_2017} and adopting the necessary privacy constraints. Our solution focuses on the SSMUPIR setting and its extension, the MUPIR setting. The fundamental process behind our scheme is as follows:
\begin{enumerate}
    \item Generate a query scheme with other users to access information from an arbitrary number of databases
    \item Send queries to the database(s) and other users
    \item Receive and combine answers from the database(s) and other users
\end{enumerate}
A visual model of the mirrored scheme we create is shown in Fig. 1. 
\begin{figure}[ht]\label{new}
    \centering
    \includegraphics[trim={0.0cm 0.7cm 0.1cm 4.2cm},clip,width=3.0in]{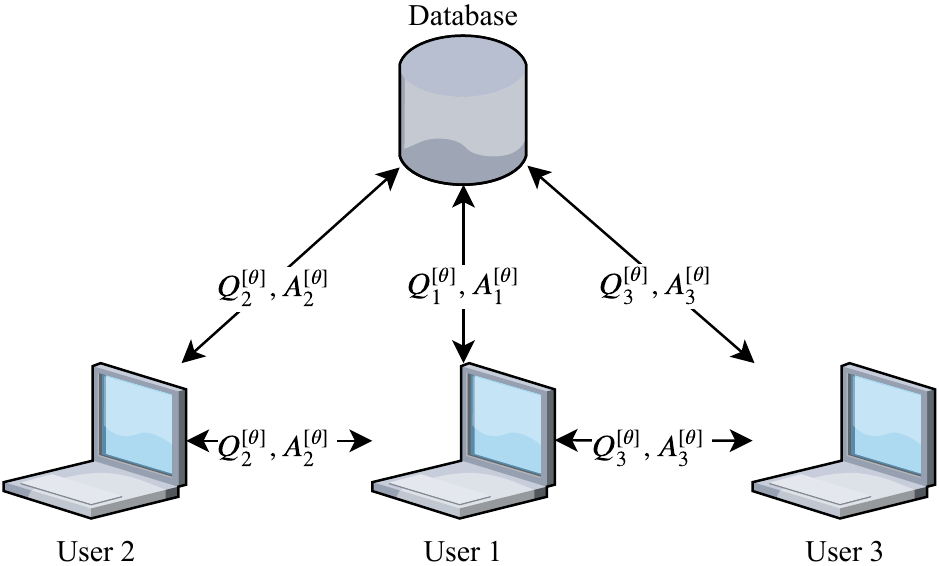}
    \caption{A visual depiction of a user attempting PIR with two other users and a database.}
\end{figure} The upper bound for capacity in a setting with a single user is achieved by a scheme utilized in \cite{sun_capacity_2017}. Calling function $Q(s,\theta)$ results in "$k$-sums," generated by the equation $\sum^{K}_{k=1}(S-1)^{k-1}\binom{K}{k}$.
There are also two other options for the $U-1$ users that are considered: In some contexts, all $U-1$ users must replicate all contents stored within a select database in order for a central user to preserve privacy from other users. This is necessary when a database cross-references queries made by multiple users and wants to see if there are any colluding users and reveal a potential desired message index. If a database is computationally bounded and cannot analyze queries made by all users, then there is no need for $U-1$ users to replicate the entirety of the database with their own means of storage. Our scheme would have to be limited in application to settings with databases that cannot afford to cross-reference queries from all users. 
We create another example in Fig. \ref{demo} while assuming the three defined principles are followed. Consider a simple PIR setting with a single database $(N=1)$ and two users $(U=2)$. Our scheme begins by having an arbitrary user (U1) request a portion of a desired message bit of $W_1$ from the database DB1, which in this case is $a_1$. The first principle (1) is applied to result in the other user (U2) to request bit $a_2$. Applying the next principle (2) results in bits $b_1$ and $b_2$ being requested by each respective source. Principle (3) is responsible for bits $b_1$ and $b_2$ being utilized as side information and compressed in each users' queries to allow for efficiency. By randomly requesting each bit from a message to create an equal probability of all message indices being desired, privacy is preserved.
\input{Tables/Example1.tex}

\noindent In this setting, a message of size $L=4$ bits is recovered from a set of $D=6$ downloaded bits. Thus, a rate of $R=\frac{L}{D}=\frac{4}{6}=\frac{2}{3}$ is achieved and satisfies the capacity in Theorem 1. To prove that rate increases, we move forward by formally proving the relationship with $N$ databases and $U$ users with capacity.

\section{Proving Capacity}
For the original setting of SSMUPIR, our solution stems from setting $N=1$ and $S=U$ in eqn. \ref{onencap},  yielding the capacity for SSMUPIR. We formally prove the capacities of SSMUPIR and MUPIR by finding the quantity of desired information, $H(W_{\theta})$ and the cardinality of $Q(s,\theta)$. We apply this to the definition of rate in (\ref{rate}) in order to find capacity.



\noindent\textit{Lemma 1:} {$\forall{S} \in \mathbb{N}_{>0}$ and $\theta \in [K]$, message $W_{\theta}$ can be successfully retrieved while maintaining privacy from all sources and achieve $R\leq(1+\frac{1}{S}+\cdots+\frac{1}{S^{K-1}})^{-1}$.}
\begin{IEEEproof}
For simplicity, let $\theta=1$ and $K=1$, to obtain message $W_{1}$. Building off the previous Lemma, we characterize the capacity of MUPIR with respect to preserving privacy from other users, utilizing the function $I(\p{A}{1}{1:U};W_{1})$. Similar to the the setting for SSMUPIR, the identity for $H(\p{A}{1}{n}|\Q) \neq H(\p{A}{1}{u}|\Q)$ applies to settings where both $U$ and $N$ are greater than 1, which is characterized by $I(\p{A}{1}{1:U};\p{A}{1}{1:N};W_{1})$ for solving capacity.\footnote{A brief overview of how we utilize multivariate mutual information is included in Appendix C.} We evaluate the upper bound of mutual information to be:
\input{Lemmas/3.tex}
\end{IEEEproof}

\section{Achievability of the Proposed Scheme}


\noindent\textit{Lemma 2:} {$\forall{\theta} \in [K]$, and $s \in [S]$, the structure of $Q(u,\theta)$ achieves the cardinality $|Q(s,\theta)|=S^{K-1}+\frac{1}{S-1}(S^{K-1})$ in a single database setting. The deterministic function can be extended to utilize settings where $N>1$ by generating $(N+U-2)^{k-1}$ instances for each $k$-sum.}

\begin{IEEEproof}
\input{Lemmas/5.tex}
\end{IEEEproof}

With the length of the desired message and cardinality of downloaded information known, we can determine the information-theoretic rate.\

\noindent\textit{Lemma 3:} {The rate achieved is equal to the capacity, $C_{\text{MUPIR}}=(1+\frac{1}{S}+\cdots+\frac{1}{S^{K-1}})^{-1}$.}
\begin{IEEEproof}
Assuming the cardinality of $Q(s,\theta)$ presented in Lemma 5 is correct, the result can applied to the definition of rate to find that the scheme is capacity-achieving.
\input{Lemmas/6.tex}
\end{IEEEproof}
\section{Conclusion}
Our adaptation of the original scheme presented in \cite{sun_capacity_2017} shows that there is a great deal of potential for undermining systems of colluding databases by introducing other users as potential sources of information. We are not surprised that the capacity of the three possible scenarios in focus is identical. Intuitively, we would expect there to be a lower capacity in other scenarios where there are malicious users or databases having knowledge of user collusion. We have successfully contributed to the field of PIR by 1) finding the upper bound of capacity for settings with multiple users in a single database setting, 2) characterized the capacity for a user for settings with a user interacting with other users alongside multiple databases, and 3) creating a PIR scheme that can be utilized in all of these settings. However, this scheme is only guaranteed to protect privacy if the database analyzes queries made by a single user at a time due to computational costs. If the database in question cross-references all queries made by all users, then privacy is at risk of being compromised. By solving this issue, a breakthrough can be made in PIR for securing privacy against many networked databases.

\input{appendix.tex}


\bibliographystyle{IEEEtran}
\bibliography{MUPIR.bib}
\end{document}

%% file: setup.tex
\usepackage{cite}
\usepackage[T1]{fontenc} 
\usepackage{mathtools,amsmath}
\usepackage[cmintegrals]{newtxmath}
\usepackage{bm} 
\usepackage{tikz}
\usetikzlibrary{er,positioning,shapes.symbols}
\usepackage{graphicx}
\usepackage{textcomp}
\usepackage{xcolor}
\usepackage{multirow}
\usepackage{comment}
\usepackage{arydshln}
\usepackage{ragged2e}
\usepackage{enumitem}
\usepackage{multicol}
\usepackage{wrapfig}

\graphicspath{ {./Images/} }

\makeatletter

\usepackage{ifpdf}
\ifpdf
\else
\fi

\usepackage{algorithm,algcompatible,float}
\usepackage{lipsum}

\makeatletter
\newenvironment{breakablealgorithm}
  {
   \begin{center}
     \refstepcounter{algorithm}
     \hrule height.8pt depth0pt \kern2pt
     \renewcommand{\caption}[2][\relax]{
       {\raggedright\textbf{\ALG@name~\thealgorithm} ##2\par}%
       \ifx\relax##1\relax 
         \addcontentsline{loa}{algorithm}{\protect\numberline{\thealgorithm}##2}%
       \else 
         \addcontentsline{loa}{algorithm}{\protect\numberline{\thealgorithm}##1}%
       \fi
       \kern2pt\hrule\kern2pt
     }
  }{
     \kern2pt\hrule\relax
   \end{center}
  }
\makeatother

\algnewcommand\algorithmicreturn{\textbf{return}}
\algnewcommand\RETURN{\State \algorithmicreturn}%

\newcommand{\p}[3]{#1^{[#2]}_{#3}}
\newcommand{\Q}{\textstyle \sum{Q}}


%% file: Tables/Example1.tex
\begin{table}[ht]
\centering
\begin{tabular}{|c|c|}
 \hline
 \tiny{U1} & \tiny{U2}\\
 \hline
 $a_{1}$ & \\
 \hline
\end{tabular}
$\xrightarrow[]{(1)}$
\begin{tabular}{ |c|c| }
 \hline
 \tiny{U1} & \tiny{U2}\\
 \hline
 $a_{1}$ & $a_{2}$ \\
 \hline
\end{tabular}
$\xrightarrow[]{(2)}$
\begin{tabular}{ |c|c| }
 \hline
 \tiny{U1} & \tiny{U2}\\
 \hline
 $a_{1},b_{1}$ & $a_{2},b_{2}$ \\
 \hline
\end{tabular}
$\cdots$
\vspace{1mm}\\
$\cdots\xrightarrow[]{(3)}$
\begin{tabular}{ |c|c| }
 \hline
 \tiny{U1} & \tiny{U2}\\
 \hline
 $a_{1},b_{1}$ & $a_{2},b_{2}$ \\
 $a_{3}+b_{2}$ & \\ 
 \hline
\end{tabular}
$\xrightarrow[]{(1)}$
\begin{tabular}{ |c|c| }
 \hline
 \tiny{U1} & \tiny{U2}\\
 \hline
 $a_{1},b_{1}$ & $a_{2},b_{2}$ \\
 $a_{3}+b_{2}$ & $a_{4}+b_{1}$ \\ 
 \hline
\end{tabular}
\end{table}
\begin{figure}[ht]
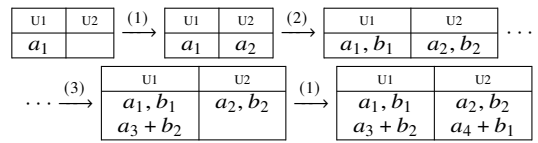

    \vspace{-20pt}
    \centering
    \caption{Demonstration of Query Generation for $K=2$, $N=1$, and $U=2$.}
    \label{demo}
    \vspace{10pt}
\end{figure}


%% file: Lemmas/3.tex
\begin{equation}
    I(\p{A}{1}{1:U};\p{A}{1}{1:N};W_{1}|\Q) \leq  I(\p{A}{1}{1:U};\p{A}{1}{1:N}|\Q)
\end{equation}
Evaluating the boundary of mutual information is critical for solving capacity. Typically, for a setting of $U$ users interfacing with a single database,  $I(\p{A}{1}{1},\p{A}{1}{2:U};W_{1}|\Q)$ could be used to evaluate how much information is downloaded. The combination of $U$ users and $N$ databases requires a different approach to be taken with an allowance for more random variables. For the case of a single user and a single database, the quantity of downloaded information is determined by $I(\p{A}{1}{1};\p{A}{1}{1};W_{1}|Q)\leq H(\p{A}{1}{1}|\Q)$. This is the result of only one source of new information, rather than two sources, due to a single user being unable to generate new information independently. This distinction causes us to substitute $I(\p{A}{1}{1:U};\p{A}{1}{1:N}|\Q)$ with $I(\p{A}{1}{2:U};\p{A}{1}{1:N}|\Q)$. In order to find our desired result, we must minimize conditional entropy so that $H(\p{A}{1}{2:U}|\p{A}{1}{1:N})=H(\p{A}{1}{1:N}|\p{A}{1}{2:U}) = 0$ to achieve
\begin{equation}
    I(\p{A}{1}{2:U};\p{A}{1}{1:N}|\Q)=H(\p{A}{1}{2:U},\p{A}{1}{1:N}|\Q)
\end{equation}
We combine answers $\p{A}{1}{2:U}$ and $\p{A}{1}{1:N}$ to create $\p{A}{1}{1:S}$ to represent new information from an equivalent quantity of $S$ sources. We conveniently characterize the new capacity starting from
\begin{equation}
    H(W_{1})=H(W_{1}|\Q)\leq I(\p{A}{1}{1:S};W_{1}|\Q)
\end{equation}
From here, we can apply the same techniques for entropy conditioning used in \cite{sun_capacity_2016} by evaluating $I(\p{A}{1}{1},\p{A}{1:K}{2:S};W_{1:K}|\Q)$ with the assumption that $K=3$. By induction, we
obtain our capacity for the MUPIR with preservation of privacy from all $S=N+U-1$ sources:
    \begin{equation}
    R \leq \Big(1+\frac{1}{S}+\cdots+\frac{1}{S^{K-1}}\Big)^{-1}
\end{equation}
This concludes our proof for capacity of all PIR settings with multiple users or databases.

%% file: Lemmas/5.tex
The query scheme used to address this problem is nearly identical to the $k$-sum equation presented in \cite{sun_capacity_2017}. A necessary modification is that $(U-1)^{k-1}$ instances for each $k$-sum must increase for each database present to generate $(N+U-2)^{k-1}$ instances. 
\begin{IEEEeqnarray}{LL}
    \textstyle|Q(u,\theta)|=\textstyle |Q(s,\theta)|&=\sum_{k=1}^{K}\binom{K}{k}(S-1)^{k-1}\\
    &=\textstyle S^{K-1}+\textstyle\frac{1}{S-1}(S^{K-1}-1) \label{newmessize}
\end{IEEEeqnarray}
We now know the relationship between downloaded information and replicated databases and can determine rate.

%% file: Lemmas/6.tex
\begin{IEEEeqnarray}{LL}
    R\triangleq \frac{L}{D}&=\frac{S^K}{S|Q(s,\theta)|}=\frac{S^{K-1}}{\sum_{k=1}^{K}\binom{K}{k}(S-1)^{k-1}}\label{capproof}\label{look}\\
    &=\bigg(1+\frac{\frac{1}{S}(1-\frac{1}{S^{K-1}})}{1-\frac{1}{S}}\bigg)^{-1}{=}\:\bigg(1+\frac{1}{S}+\cdots+\frac{1}{S^{K-1}}\bigg)^{-1}\label{done}
\end{IEEEeqnarray}
The rate achieved by (\ref{done}) is equal to the capacity presented in Theorem 1.

%% file: appendix.tex
\appendices
\section{Algorithm for Brute Force Analysis of a Query Made by an Individual User to Find the Desired Message Index}
\input{Algorithms/1}
\section{Algorithm for Brute Force Analysis of Queries Made by Multiple Users to Find the Desired Message Index}
\input{Algorithms/2}
\section{Characterization of Multivariate Mutual Information}
To characterize $I(\p{A}{1}{1:U};\p{A}{1}{1:N};W_{1})$, we apply the following: From Theorem 2 of \cite{yeung_new_1991}, the upper bound for mutual information utilizing three random variables ($X$, $Y$, and $Z$) is
\begin{equation}\IEEEnonumber
    I(X;Y;Z) \leq \min\{ I(X;Y),I(X;Z),I(Y;Z)\}
\end{equation}
This identity stems from using mutual information in the form $I(X;Y)$. The upper bound of interaction information can be expressed as
\begin{equation}\IEEEnonumber
    H(W_{1}) \leq \min\{ I(\p{A}{1}{1:U};\p{A}{1}{1:N}),I(\p{A}{1}{1:U};W_{1}),I(\p{A}{1}{1:N};W_{1})\}
\end{equation}
By definition, $I(\p{A}{1}{1:U};W_{1})$ and $I(\p{A}{1}{1:N};W_{1})$ must be maximized in order to achieve capacity, thus obtaining:
\begin{equation}\IEEEnonumber
     H(W_{1}) \leq  I(\p{A}{1}{1:U};\p{A}{1}{1:N})
\end{equation}

%% file: Algorithms/1.tex
  \begin{breakablealgorithm}
  \caption{Input: Query Set $Q(\text{User}, \theta)$. Output: $\theta$.}
  \begin{algorithmic}[1]
        \REQUIRE{All queries from a distinct user to a database are complete.}
    \STATE Initialize: Associative array $\mu$ containing all bits for $K$ messages as keys with their respective indices mapped as values $\forall{k} \in [K]$. Initialize array $\beta$ of length $K$ containing all zeroes.
    \FORALL[For each bit...]{$\alpha \in Q(\text{User},\theta)$}
    \IF[...if a bit is unidentifiable...]{$\alpha \notin \mu$}
        \FORALL[...then compare to all other bits...]{$\alpha'\in Q(\text{User},\theta)$}
            \IF[...if an identifiable bit is generated...]{($\alpha$ $\oplus$ $\alpha'$) $\in \mu$}
               \STATE{$\beta$[$\mu$[$\alpha$ $\oplus$ $\alpha'$]] += 1}
               \COMMENT{...increment the respective index in the list for storing the numbers of identified bits.}
            \ENDIF
        \ENDFOR
    \ELSE[If the bit is already distinguishable...]
        \STATE{$\beta$[$\mu$[$\alpha$]] += 1}
        \COMMENT{...increment the respective index...}
    \ENDIF
    \ENDFOR
    \RETURN{ max($\beta$)}
    \COMMENT{...and end by returning $\theta$.}
  \end{algorithmic}
  \end{breakablealgorithm}

%% file: Algorithms/2.tex
  \begin{breakablealgorithm}
  \caption{Input: Query Sets $\Q$. Output: $\theta$.}
  \begin{algorithmic}[1]
    \REQUIRE{All queries from a distinct set of users to a database are complete.}
    \STATE{Initialize: Associative array $\mu$ containing all bits for $K$ messages as keys with their respective indices mapped as values $\forall{k} \in [K]$. Initialize array $\beta$ of length $K$ containing all zeroes.}
    \FORALL[For each user...]{$u \in [U]$}
        \FORALL[...analyze each bit...]{$\alpha \in Q(u,\theta)$}
            \FORALL[...by comparing to other users...]{$u' \in [U']$}
            \IF[...if a bit is unidentifiable...]{$\alpha \notin \mu$}
                \FORALL[...compared to all other bits...]{$\alpha'\in Q(u',\theta)$}
                    \IF[...if an identifiable bit is generated...]{($\alpha$ $\oplus$ $\alpha'$) $\in \mu$}
                       \STATE{$\beta$[$\mu$[$\alpha$ $\oplus$ $\alpha'$]] += 1}
                       \COMMENT{...update the respective index count in the list for storing the numbers of identified bits.}
                    \newpage
                    \ENDIF
                \ENDFOR
            \ELSE[If the bit is already distinguishable...]
                \STATE{$\beta$[$\mu$[$\alpha$]] += 1}
                \COMMENT{...increment the respective index...}
            \ENDIF
            \ENDFOR
        \ENDFOR
    \ENDFOR
    \RETURN{ max($\beta$)}
    \COMMENT{...and return $\theta$.}
  \end{algorithmic}
\end{breakablealgorithm}